\documentclass[reprint,superscriptaddress,amsmath,amssymb,amsart,aps,prb]{revtex4-2}
\usepackage{accents}
\usepackage{graphicx}
\usepackage{dcolumn}
\usepackage{bm}
\usepackage{braket}
\usepackage{amsthm}
\usepackage{bookmark} 
\usepackage[autostyle]{csquotes}
\usepackage{hyperref}
\usepackage{booktabs}
\graphicspath{{./figs/}}
\usepackage{lineno}

\begin{document}

\title{Quantum Kinetic Anatomy of Electron Angular Momenta Edge Accumulation}

\author{Thierry Valet}
\email[Corresponding author :]{ tvalet@mphysx.com}
\affiliation{MPhysX O\"U, Harju maakond, Tallinn, Lasnam\"ae linnaosa, Sepapaja tn 6, 15551, Estonia}

\author{Henri Jaffr\`es}
\affiliation{Laboratoire Albert Fert, CNRS, Thales, Universit\'e Paris-Saclay, 91767 Palaiseau, France}

\author{Vincent Cros}
\affiliation{Laboratoire Albert Fert, CNRS, Thales, Universit\'e Paris-Saclay, 91767 Palaiseau, France}

\author{Roberto Raimondi}
\affiliation{Dipartimento di Matematica e Fisica,  Universit\`a  Roma Tre, Via della Vasca Navale 84, 00146 Roma, Italy}

\date{\today}

\begin{abstract}
Controlling electron's spin and orbital degrees of freedom has been a major research focus over the past two decades, as it underpins the electrical manipulation of magnetization. Leveraging a recently introduced quantum kinetic theory of multiband systems [T. Valet and R. Raimondi,
{\em Phys. Rev. B} 111, L041118 (2025)], we outline how the intrinsic angular momenta linear response is partitioned into intraband and interband contributions. Focusing on time reversal and inversion symmetric metals, we show that the spin and orbital Hall currents are purely intraband. We also reveal that the intrinsic edge densities originate partially, and in the orbital case probably mostly, from a new interband mechanism. We discuss how this profoundly impacts the interpretation of orbital edge accumulation observations, and has broader implications for current induced torques.    
\end{abstract}

\maketitle

\setlength\linenumbersep{5pt}
 
Electrical generation of out-of-equilibrium spin and orbital electronic angular momenta (AMs) in nonmagnetic materials are central phenomena in the fields of spin-orbitronics \cite{sinova:2015, soumya:2016} and orbitronics \cite{bernevig:2005, go:2021}, respectively. The growing interest and rapid progress in these areas are largely driven by their potential to manipulate magnetization \cite[\& references therein]{manchon:2019,rappoport:2023, Lee:2024, atencia:2024,jo:2024}. The intrinsic AMs transverse current response in time reversal (TR) and inversion (I) symmetric metals continue to be a central focus. Early work concentrated on heavy \emph{5d} metals ({\em e.g.}, Pt, Ta or W), where strong spin-orbit interaction (SOI) gives rise to a large, theoretically predicted, intrinsic spin Hall effect (SHE) \cite{guo:2008,qiao:2018}. More recently, interest has shifted to light elements with weak SOI ({\em e.g.}, Ti, Cr, V or Mn), spurred by predictions of a ``giant'' intrinsic orbital Hall effect (OHE), remarkably found independent of the SOI \cite{tanaka:2008,kontani:2009,go:2018,go:2018b,pezo:2022,busch:2023}. Observations of spin \cite{stamm:2017,pattabi:2023} and orbital \cite{choi:2023,lyalin:2023,idrobo:2024} edge accumulation, at the boundaries of thin-film elemental-metal conductors are widely regarded as the most direct experimental confirmations of the predicted intrinsic SHE and OHE, respectively. This interpretation relies on analysis assuming diffusive transport of the electronic AMs, in which the intrinsic SHE or OHE are treated as source terms \cite{stamm:2017,lyalin:2023}. 

\begin{figure}[ht]
\includegraphics[width=0.34\textwidth]{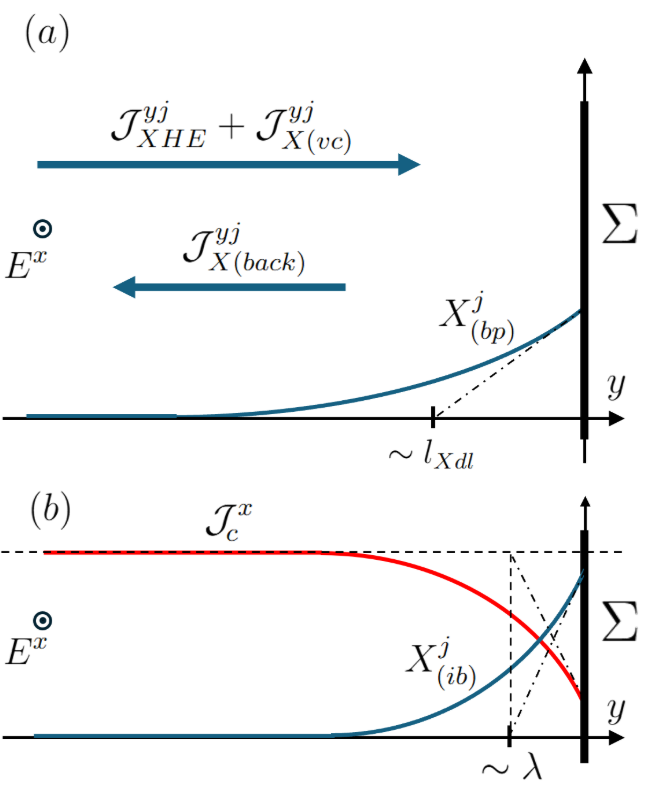}
\vspace{-10 pt}
\caption{\label{fig:1} Near a boundary $\Sigma$ of a (TR+I) symmetric metal conductor, a tangential electric field \smash{$E^x$} is inducing a density of X, with X standing for either spin (S) or orbital (L) angular momentum. (a) The band projected part \smash{$X^{j}_{(bp)}$}, is confined within a few angular momentum diffusion lengths \smash{$l_{Xdl}$}, resulting from the detailed balance between the intrinsic angular momentum Hall current with vertex correction \smash{$\mathcal{\bm J}^{y j}_{XHE} + \mathcal{\bm J}^{y j}_{X (vc)}$,} the backflow current \smash{$\mathcal{\bm J}^{y j}_{X (back)}$} and the angular momemtum relaxation. (b) The interband part \smash{$X^{j}_{(ib)}$}, is confined within a few electron mean free paths \smash{$\lambda$}. It results from quantum coherences induced by the transverse gradient in the longitudinal response, which also manifests itself by a decreased charge current density \smash{$\mathcal{J}_c^x$} near the edge.}
\vspace{-10 pt}
\end{figure}

In this Letter, we present a unified quantum kinetic theory of the intrinsic spin and orbital Hall effects and AMs edge accumulations in metals with (TR+I) symmetry. Departing from previous theoretical approaches, we demonstrate that the physically relevant Hall effects corresponds to the transverse currents of the {\em band projected} AMs. Except in (rare) cases of orbital degeneracy \cite{bernevig:2005}, this leads to the conclusion that the intrinsic OHE is second order in the SOI. As another key result, we uncover a novel contribution to AMs edge accumulation stemming from gradients in the intraband longitudinal response. These gradients {\em locally} generate interband coherences -- a general effect  recently predicted by Valet and Raimondi \cite{valet:2025} -- and naturally arise near conductor boundaries \cite{fuchs:1938,sondheimer:1952,camley:1989}. A concrete calculation for a simple model of (TR+I) symmetric metals \cite{tokatly:2010,han:2022,tang:2024} reveals it as a manifestation of {\em quantum orbital AM - vorticity coupling}. While this new effect likely dominates the intrinsic orbital edge accumulation in non-magnetic centrosymmetric metals with weak SOI, it has broader implications for our understanding of current induced torques, opening new perspectives for the fields of spin-orbitronics \mbox{and orbitronics.} 

We consider a generic crystal exhibiting (TR+I) symmetry, whose electronic structure in the vicinity of the Fermi level is assumed to be well approximated by a low energy Hamiltonian \smash{$\hat{h}(\bm p) = \sum_n \varepsilon_n \hat{P}_n$}, with ${\bm p}$ the kinetic pseudo-momentum, $n$ the band index, \smash{$\varepsilon_n$} the band energies and \smash{$\hat{P}_n$} the associated eigenprojectors \cite{graf:2021,mitscherling:2024}. Excluding any rare orbital degeneracies beyond Kramers’ theorem, we focus on the generic case in which each band is twofold degenerate under the assumed symmetry. We introduce the single-particle observable \smash{$\hat{\bm X} \equiv \hat{X}^j$,} with the index $j$ labeling its different spatial components. This observable stands for either the spin \smash{($\hat{\bm S}$)} or the orbital \smash{($\hat{\bm L}$)} AM operator, as appropriate. As for any other operator in the considered Hilbert space, we can perform a {\em unique} decomposition as \smash{$\hat{\bm X} = \hat{\bm X}_{(bp)} + \hat{\bm X}_{(ib)}$}, where \smash{$\hat{\bm X}_{(bp)} = \sum_n \hat{P}_n \hat{\bm X} \hat{P}_n = \sum_n \hat{\bm X}_n$} is the band-projected part, and \smash{$\hat{\bm X}_{(ib)} = \sum_{n \neq m} \hat{P}_n \hat{\bm X} \hat{P}_m  = \sum_{n \neq m} \hat{\bm X}_{n m}$} carries the interband contributions. This decomposition follows directly from the completeness relation of the projectors. We are interested in the expectation value \smash{${\bm X} \equiv \langle \hat{\bm X} \rangle$,} evaluated in linear response to an electric field ${\bm E}$ which we assume to be quasi-uniform and slowly varying in time. Building on our recent quantum kinetic theory of multiband systems \cite{valet:2023, valet:2025}, and as detailed in the Supplemental Material \cite{supp}, we obtain a continuity equation for the band-projected part
\begin{equation} \label{eq:curr_cont}
    \partial_t X^{j}_{(bp)} + \partial_{i} {\mathcal J}^{i j}_{X} =  \partial_t X^{j}_{(bp)} \!  \left. \right|_{(relax)} ,
\end{equation}
which {\em defines} the AM current \smash{${\mathcal J}^{i j}_{X}$}, while the right-hand-side (RHS) captures the disorder induced AM relaxation processes. As for the interband part, it is determined by the field induced quantum coherences. It reads
\begin{equation} \label{eq:x_inter}
    {\bm X}_{(ib)} = \sum_{n \neq m} \  \int_{p} {\rm Tr} \left[ \hat{\rho}_{n m}  \hat{\bm X}_{m n} \right] ,
\end{equation}
with \smash{$\int_{p} \equiv \int\displaylimits_{\scriptscriptstyle BZ} \frac{d^d p}{(2 \pi \hbar)^d}$}   defining the momentum integration over the Brillouin zone (BZ) of dimension $d$. A general expression of the band off-diagonal elements of the density matrix, {\em i.e.},  \smash{$\hat{\rho}_{n m}$}, has been derived in \cite{valet:2025}, where it is thoroughly established that these interband coherences are {\em locally} induced quantities -- and consequently, so is \smash{${\bm X}_{(ib)}$}. This local character implies that {\em no long range AM current} is associated with the interband contribution. Therefore, the physical AM current can be identified with \smash{${\mathcal J}^{i j}_{X}$} as defined by Eq.(\ref{eq:curr_cont}). It comprises three distinct contributions 
\begin{equation} \label{eq:x_curr}
 {\mathcal J}^{i j}_{X} = {\mathcal J}^{i j}_{XHE} + {\mathcal J}^{i j}_{X(vc)} +{\mathcal J}^{i j}_{X (back)} .
\end{equation}
The first term is the {\em intrinsic}, spin or orbital, Hall current
\begin{equation}\label{eq:xhe_curr}
    {\mathcal J}^{i j}_{XHE} = -e \hbar \epsilon_{i k l}  E^k  \sum_n \!  \int_{p} n_{\scriptscriptstyle FD}(\varepsilon_{n}) {\rm Tr} \left[ \hat{X}^j_n  \hat{\mathcal F}^l_n \right] , \!\!\! 
\end{equation}
with \smash{$\epsilon_{i k l}$} the Levi-Civita symbol, and with the integration being carried over the occupied states in equilibrium, as determined by the Fermi-Dirac distribution \smash{$n_{\scriptscriptstyle FD}$}. Hence, we find the intrinsic SHE and OHE to be governed by the same pseudovector Berry curvature  \smash{$\hat{\mathcal F}_{n}^l = ( \hat{\bm{\mathcal F}}_{n} )^{l} = (1/2) \epsilon_{l a b} \hat{\mathcal F}^{a b}_{n}$,} with \smash{$\hat{\mathcal F}^{a b}_{n} = {\bf i} \hat{P}_n [ \partial_p^a \hat{P}_n, \partial_p^b \hat{P}_n ]_{\scriptscriptstyle (-)} \hat{P}_n$,} the non-Abelian Berry curvature tensor and \smash{$ \left[ \cdot , \cdot \right]_{\scriptscriptstyle (-)}$} denoting a commutator. While an equivalent expression of the intrinsic SHE has been previously obtained using the wave-packet formalism \cite{shindou:2005,xiao:2020}, our derivation provides for the first time a rigorous formulation unifying the SHE and OHE, from a controlled semiclassical expansion of the Keldysh theory. Furthermore, it is to be emphasized that most recent theoretical studies incorrectly compute these effects from the expectation value, taken from the Fermi sea term of the Kubo formula, of the conventional AMs current operators, {\em i.e.}, \smash{$\hat{\mathcal J}^{i j}_{X (conv)} = \frac{1}{2} [ \partial_{p^i} \hat{h} , \hat{X}^j ]_{\scriptscriptstyle (+)}$}, in which \smash{$ \left[ \cdot , \cdot \right]_{\scriptscriptstyle (+)}$} denotes an anti-commutator. This seems to be solely based on an unfounded analogy with the Kubo formula for the intrinsic anomalous Hall effect (AHE), and lacks rigorous justification. It leads to the introduction of the so-called spin and orbital Berry curvatures, which are nonphysical and shall not be mistaken for the non-Abelian Berry curvature \smash{$\hat{\bm{\mathcal F}}_{n}$} truly controlling the SHE and OHE.  Interestingly, the continued use of this method persists despite the fact that the seminal works of Murakami {\em et al} \cite{murakami:2004b} on SHE theory, and of Bernevig {\em et al} \cite{bernevig:2005} on OHE theory, had already outlined two decades ago that the conventional AMs current operators are not physical. The alternative definition of conserved current operators proposed in \cite{murakami:2004b,bernevig:2005}, given by the expectation value of \smash{$\hat{\mathcal J}^{i j}_{X (cons)} = \frac{1}{2} [ \partial_{p^i} \hat{h} , \hat{X}^j_{(bp)} ]_{\scriptscriptstyle (+)}$}, while an appealing concept, still lacks a rigorous justification and does not generally lead to a correct evaluation of the intrinsic SHE and OHE currents as given by our Eq.(\ref{eq:xhe_curr}). In the spin case, and although different definitions of the current can lead to quantitative differences in  predicted values of the intrinsic SHE conductivity \cite{gradhand:2012}, the qualitative conclusion that the intrinsic SHE is first order in the SOI strength is unaffected. In contrast, for the orbital case, an appropriate definition of the current is crucial. The conventional current choice leads to the erroneous conclusion that the intrinsic OHE in light transition metals is independent of the SOI. Our Eq.(\ref{eq:xhe_curr}) establishes that in (TR+I) symmetric metals, and in the absence of exceptional orbital degeneracy, the intrinsic OHE is at least second order in the SOI \cite{endmatter}. This is because both the intraband part of the orbital AM operator and the Berry curvature vanish in the zero SOI limit under the stated hypothesis \cite{supp}. Finally, even when focusing on intrinsic effects, one must account for the potential contribution of disorder induced vertex corrections, which corresponds to the second term in Eq.(\ref{eq:x_curr}). While such corrections have been considered before in the contexts of the intrinsic SHE and OHE, with at times dramatic effects \cite{inoue:2004, mishchenko:2004,raimondi:2005,tang:2024}, these analyses were based on the conventional current operator definition. Although this is beyond the scope of the present work, this will have to be analyzed anew within our quantum kinetic framework, with connections to be clarified with the recently introduced concept of {\em extrinsic velocity} \cite{atencia:2022}, as outlined in the Supplemental Material \cite{supp}.  

Switching our attention to the edge accumulation of AMs, we consider now a finite system, with an electric field applied parallel to one of its boundaries $\Sigma$, as illustrated in Fig.(\ref{fig:1}). This configuration generally gives rise to a non-zero steady state AM density \smash{$X^{j}$} in the vicinity of $\Sigma$. We first focus on its band projected part, whose spatial gradient is linked to the backflow current \smash{${\mathcal J}^{i j}_{X (back)}$,} {\em i.e.}, the third term on the RHS of Eq.(\ref{eq:x_curr}). To obtain a closed formulation for \smash{$X^{j}_{(bp)}$}, the following elements are needed : (i) an explicit form for the relaxation term on the RHS of Eq.(\ref{eq:curr_cont}), for instance \smash{$\partial_t X^{j}_{(bp)}|_{(relax)} = -({\bar{\bar\tau}}_X^{-1})^j{}_i X^{i}_{(bp)}$,} with \smash{${\bar{\bar\tau}}_X$} the AM relaxation time tensor, (ii) the determination of the AM diffusion tensor $\bar{\bar{\mathcal D}}_X$ so that \smash{${\mathcal J}^{i j}_{X (back)} = -{\mathcal D}_X^{i j k}{}_l \partial_k X^{l}_{(bp)}$,}
and (iii) the specification of the boundary conditions on $\Sigma$, for instance with the introduction of an AM relaxation velocity tensor \smash{$\bar{\bar{\mathcal V}}$,} leading to \smash{$ n_i {\mathcal J}^{i j}_{X}|_{\Sigma} = {\mathcal V}^{i j}{}_k X^{k}_{(bp)}|_{\Sigma}$,} with \smash{${\bm n} \equiv n^i$} the local outgoing normal vector at the boundary. Such parameterization allows for a varying degree of conservation of the AM at the boundary. A vanishing \smash{$\bar{\bar{\mathcal V}}$} corresponds to perfect AM conservation and maximum edge accumulation, while an infinite relaxation velocity corresponds to perfect absorption of the incoming AM current and vanishing edge accumulation. Assuming isotropic relaxation and diffusion for simplicity, one can derive (iv) the diffusion equation \smash{$\Delta X^{i}_{(bp)} = X^{i}_{(bp)} / l_{Xdl}^2$,} with $\Delta$ the Laplacian and \smash{$l_{Xdl} = \sqrt{{\mathcal D}_X \tau_X}$} the AM diffusion length that sets the spatial extent of the band-projected part of the edge accumulation. This diffusive transport framework aligns with the common models used to analyze recent experiments, and is illustrated in Fig.\ref{fig:1}(a). While our quantum kinetic framework is ideally suited for deriving the transport parameters introduced in (i-iv), we choose not doing it here, as this lies within the scope of established phenomenology. Importantly, as we have shown, the Hall effect related AM edge accumulation only captures the band projected contribution.

We identify now a novel contribution to the edge accumulation which is confined within a few electron mean free paths, denoted  \smash{$\lambda$,} and which arises from the interband part of the AM density. As well established in the Fuchs-Sondheimer theory \cite{fuchs:1938,sondheimer:1952} of the resistivity of thin metal films, linear momentum non-conserving scattering processes at conductor boundaries, associated with atomic-scale surface roughness, leads to a transverse gradient in the longitudinal response to an electric field \cite{endmatter}. This is confined in a narrow region at the conductor edges scaling with the MFP. Within this kinetic boundary layer, the longitudinal charge current density exhibits a quasi-exponential variation, reaching a value at the surface that is significantly reduced compared to the bulk, as shown in Fig.\ref{fig:1}(b). This is quantitatively captured by the Boltzmann equation controlling the longitudinal linear response in terms of \smash{$f_n$,} the deviation of the electron distribution function in band $n$. It reads
\begin{equation} \label{eq:boltz0}
    \partial_{\bm p} \varepsilon_{n} \cdot \partial_{\bm x} f_n + (e {\bm E}) \cdot \partial_{\bm p} n_{\scriptscriptstyle FD}(\varepsilon_{n}) =  \mathcal{C}_n \left[ {\bm f} \right] ,
\end{equation}
with the bulk disorder collision integral on the RHS. 

Recently, Valet and Raimondi\cite{valet:2025} have shown that such a spatial gradient in \smash{$f_n$} can {\em locally} induce  quantum coherences. In the present context, this gives rise to an additional contribution to the total AM edge accumulation which, to the best of our knowledge, has not been previously considered. From Eq.(\ref{eq:x_inter}) we obtain
\begin{equation} \label{eq:x_int2}
    \!\!\!\! X^{j}_{(ib)} \! = \! - \frac{\hbar}{2} \! \sum_{\substack{n \\  m \neq n}} \! \int_{p} \! \frac{\partial f_n}{\partial x^i} {\rm Tr} \! \left[ \hat{\mathcal A}^i_{n m} \hat{X}^j_{m n} \! + \! \hat{X}^j_{n m} \hat{\mathcal A}^i_{m n}   \right] \! + \! X^{j}_{(vc)} \!, \!\!\! 
\end{equation}
in which \smash{$\hat{\mathcal A}^i_{n m} = {\bf i} \partial_p^i \hat{P}_n \hat{P}_m$} is the non-Abelian interband Berry connection, and with \smash{$X^{j}_{(vc)}$} a possible vertex correction \cite{supp}. In presence of SOI, this may induce an SHE-independent spin accumulation and may help explain some recent puzzling numerical results  \cite{belashchenko:2023}. More critically, Eq.(\ref{eq:x_int2}) defines the sole contribution to the orbital edge accumulation which remains non-vanishing as the SOI approaches zero. Indeed, in this limit, the orbital AM operator becomes purely interband \cite{supp} and the possibility of intraband orbital transport disappears.

\begin{figure}[!t]
\includegraphics[width=0.36\textwidth]{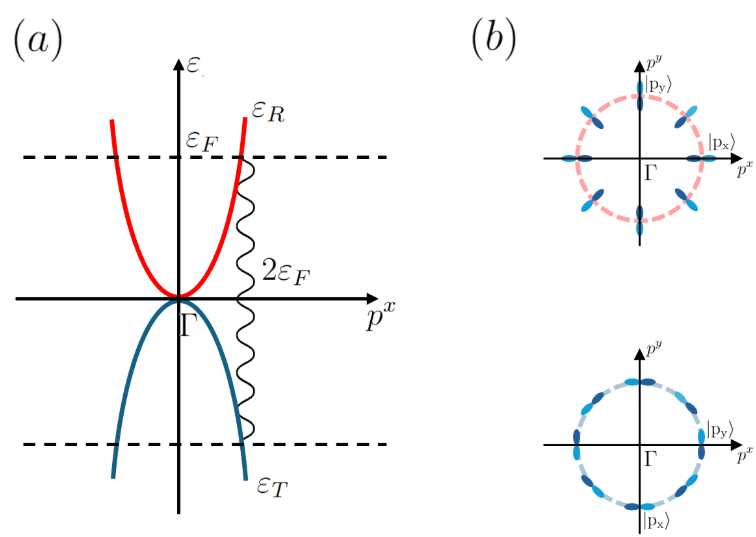}
\vspace{-5 pt}
\caption{\label{fig:2} (a) Schematic band structure of the R-T Hamiltonian model of two-dimensional (TR+I) symmetric metal considered in this work, for a concrete calculation of the interband effect. A positive Fermi level intersects only the electron like band, but virtual interband transition may occurs, see the text.  (b) P-orbital composition textures of the eigenstates on the Fermi ``surface'' in momentum space, for both bands.}
\vspace{-10 pt}
\end{figure}

To gain further insights in this newly uncovered effect, which is likely to be the dominant contribution to the orbital edge accumulation in weak SOI transition metals, we perform a concrete calculation for a simple model of two-dimensional (2D) (TR+I) symmetric metal often introduced to illustrate orbital physics {\em i.e.}, the so-called radial-tangential (R-T) \smash{$(\rm p_x,p_y)$}-orbital model \cite{tokatly:2010,go:2018,han:2022}, \cite{endmatter}. In the minimal version we choose to consider, the Hamiltonian matrix reads \smash{$ \hat{h} = \varepsilon_{\scriptscriptstyle R}\hat{P}_{\scriptscriptstyle R} + \varepsilon_{\scriptscriptstyle T} \hat{P}_{\scriptscriptstyle T}$.} As depicted in Fig.\ref{fig:2}(a), it comprises an electron like band \smash{$\varepsilon_{\scriptscriptstyle R}({\bm p}) = p^2 / (2 m^\star)$}, and a hole like band \smash{$\varepsilon_{\scriptscriptstyle T}({\bm p}) = - p^2 / (2 m^\star)$,} with \smash{$m^\star$} the effective mass, and with the pseudo-momentum ${\bm p}$ confined in the $(x, y)$ plane. We assume the Fermi level \smash{$\varepsilon_F$} to be positive, so that it only intersects the electron like band. With the momentum angle \smash{$\theta_{\bm p}$} implicitly defined by \smash{$\cos \theta_{\bm p} = p^x / p$} and \smash{$\sin \theta_{\bm p} = p^y / p$,} the electron like band corresponds to eigenstates with a radial orbital composition {\em i.e.}, \smash{$\hat{P}_{\scriptscriptstyle R} = |R \rangle \langle R |$} with \smash{$ |R \rangle = \cos \theta_{\bm p} |{\rm p}_x \rangle + \sin \theta_{\bm p} |{\rm p}_y \rangle$} ; as for the hole like band, it corresponds to eigenstates with a tangential orbital composition {\em i.e.}, \smash{$\hat{P}_{\scriptscriptstyle T} = |T \rangle \langle T |$} with \smash{$ |T \rangle = - \sin \theta_{\bm p} |{\rm p}_x \rangle + \cos \theta_{\bm p} |{\rm p}_y \rangle$,} as illustrated in Fig.\ref{fig:2}(b) on the Fermi ``surface''. Since we assume vanishing SOI, the twofold Kramers´ degeneracy of the bands reduces to a pure spin degeneracy. One can then choose a global spin‐quantization axis, allowing the spin degree of freedom to be entirely ignored. As for the orbital AM operator, adopting the atomic center approximation, we recognize that its only surviving component within the considered Hilbert space is the $z$ component, {\em i.e.,} \smash{$\hat{L}^z = \hbar \hat{\tau}^y$,} with \smash{$\hat{\tau}^y$} the $y$ Pauli matrix acting over the  \smash{$\lbrace |{\rm p}_x \rangle, |{\rm p}_y \rangle \rbrace$} orbital basis. This is to be expected for electrons whose orbital degrees of freedom are confined in the $(x,y)$ plane. We assume zero temperature and a static disorder model consisting of an homogeneous dilute random distribution of point scalar potentials, treated at the level of the second Born approximation for the derivation of the collision integral on the RHS of Eq.(\ref{eq:boltz0}). While we consider this system to be infinite in one direction in the plane, along which the electric field is applied, we assume a finite width $W$ in the perpendicular direction. As a natural choice of boundary condition at the kinetic level, we consider the edges to perfectly randomize the linear momentum of the electrons upon backscattering. The sole dimensionless parameter determining the universal scaling behavior of the longitudinal response of this system, under the stated hypothesis, is then the Knudsen number defined as the ratio between the electron MFP at the Fermi level and the conductor width, {\em i.e.}, \smash{${\rm Kn} = \lambda / W$.} In the experimentally relevant regime ${\rm Kn} \ll 1$, and as further detailed in the Supplemental Material \cite{supp}, one can derive an asymptotically accurate approximation for the electron longitudinal response, corresponding to a generalized hydrodynamic approximation  \cite{vignale:1999,tokatly:2000,struchtrup:2004b} of the electron gas kinetics. Substituting this approximate solution of the Boltzmann equation (\ref{eq:boltz0}) into Eq.(\ref{eq:x_int2}), properly specialized for the considered system, we obtain  
\begin{equation} \label{eq:Lz_hydro}
    L^z = \frac{\hbar \Omega^z}{2  \varepsilon_F} \hbar + O({\rm Kn}^3) , 
\end{equation}
with \smash{$\Omega^z = (\partial_x {\mathcal J}^y_{\! c} -  \partial_y {\mathcal J}^x_{\! c}) / e$} the {\em vorticity} associated with the electron particle current density \smash{$\bm{\mathcal J}_{\!\! c} / e$.} This equation deserves close consideration for its remarkable physical transparency. It makes manifest how the energy density associated with the vorticial part of the {\em classical} electron flow, {\em i.e.},  \smash{${\hbar \Omega^z}$}, induces virtual quantum transitions -- depicted by the wigly line in Fig.\ref{fig:2}(a) -- accross the interband energy gap $2 \varepsilon_F$ on the Fermi  ``surface'', leading to a {\em local} orbital AM density. This is made possible by the existence of a momentum dependence of the orbital composition of the eigenstates, as quantified by the interband Berry connection appearing in Eq.(\ref{eq:x_int2}). At the considered level of asymptotic approximation of the longitudinal kinetics, the charge current density obeys the following {\em generalized} electron hydrodynamic equations
\begin{subequations} \label{eq:hydro}
\begin{eqnarray} 
\partial_i {\mathcal J}^i_{\! c} & = & 0 , \label{eq:hydro1} \\
{\mathcal J}^i_{\! c} = \sigma E^i &+& (\lambda / \sqrt{2})^2 \ \Delta {\mathcal J}^i_{\! c} ,  \label{eq:hydro2}
\end{eqnarray}
\end{subequations}
with $\sigma$ the conductivity, and in which \smash{$(\lambda / \sqrt{2})$} can be interpreted as the {\em vorticity} diffusion length \cite{happel:1983}, as induced here by disorder scattering. These equations need to be solved in accordance with boundary conditions (BCs) reflecting at the hydrodynamic level the assumed scattering properties of the edges \cite{maxwell:1879}. In the present case, we obtain an impenetrability condition \smash{$
{\mathcal J}^n_{\! c} |_{\scriptscriptstyle \Sigma} = 0$} and a slip boundary condition \smash{$
\partial_n {\mathcal J}^t_{\! c} |_{\scriptscriptstyle \Sigma} = - \frac{4}{3 \pi \lambda} {\mathcal J}^t_{\! c} |_{\scriptscriptstyle \Sigma}$},
with \smash{${\mathcal J}^n_{\! c}$} and \smash{${\mathcal J}^t_{\! c}$} the normal and tangential components of the current, and \smash{$\partial_n$} the normal derivative. After solving Eqs.(\ref{eq:hydro}) for the considered geometry, the current density profile reads 
\begin{equation} \label{eq:jx}
    {\mathcal J}^x_{\! c} (y) =  \left\lbrace 1 - \mathcal{B} \ \frac{\cosh \left[ \frac{y}{(\lambda/\sqrt{2})} \right]}{\cosh\left[\frac{(W/2)}{(\lambda/\sqrt{2})} \right]} \right\rbrace (\sigma E).
\end{equation}
in which it is assumed that the infinite dimension and the applied electric field are along the $x$ direction, while the origin is taken in the middle of the ribbon conductor of width $W$ in the $y$ direction. In Eq.(\ref{eq:jx}), \smash{$\mathcal{B}$} is a BCs dependent numerical factor, with $\mathcal{B} = [1 + \frac{3 \pi}{4  \sqrt{2}} \tanh ( \frac{1}{\sqrt{2} \rm K_n})]^{(-1)}$  and \smash{$\lim_{{\rm K_n} \rightarrow 0} \mathcal{B} \approx 0.375$} in the present case. Substituting Eq.(\ref{eq:jx}) into Eq.(\ref{eq:Lz_hydro}), we finally obtain
\begin{equation} \label{eq:lz}
   L^z(y) =  \frac{\hbar}{2 \varepsilon_F} \frac{\sqrt{2} E \mathcal{B}}{e \rho \lambda} \frac{\sinh \left[ \frac{y}{(\lambda/\sqrt{2})} \right]}{\cosh\left[\frac{(W/2)}{(\lambda/\sqrt{2})} \right]} \hbar ,
\end{equation}
with $\rho = 1 / \sigma$ the resistivity and with the factor $\rho \lambda$ being recognized as independent of the disorder strength. Quite remarkably, the {\em functional form} of this profile is identical to the one derived from the assumption of a diffusive orbital transport driven by OHE \cite{lyalin:2023,idrobo:2024}, and is in qualitative agreement with the experimental findings of a quasi-exponential localization of orbital AM densities of opposite signs at the two opposite edges of ribbon conductors \cite{idrobo:2024}. However, and this is the critical difference with previous analysis, we have demonstrated that this is {\em not} an evidence of an OHE-driven diffusive orbital transport. In light transition metals, the intraband orbital transport is likely to be largely suppressed, and the orbital AM to be instead {\em locally} and continuously generated by the electron current vorticity through interband quantum processes, while being continuously absorbed by the lattice, leading to a steady edge accumulation, as given by Eq.(\ref{eq:lz}) in the case of our toy model. There is {\em no} orbital diffusion in such situation, in agreement with some earlier numerical findings from ab-initio calculations \cite{rang:2024}. The orbital edge accumulation spatial extent is of the order of the mean free path as depicted in Fig.\ref{fig:1}(b), hence {\em not} scaling with a purported orbital diffusion length. This prediction appears to be in reasonable agreement with the experimentally measured orbital accumulation thickness in Ti \cite{idrobo:2024} and Cr \cite{lyalin:2023} thin films \cite{endmatter}, both of the order of $7 \ nm$, which is close to what is expected to be the mean free path value. While we have confirmed a diffusion-like profile for the edge accumulation, what is diffusing is {\em not} the orbital AM, but its {\em local} source,  {\em i.e.,} the classical electron flow vorticity.  

In conclusion, we have developed a novel quantum kinetic theory which unifies the description of the spin and orbital Hall effects, together with the concomitant edge accumulations of angular momentum, in non-magnetic centrosymmetric metals. We demonstrate that physical angular momentum currents shall be defined as intraband currents, with the intrinsic spin and orbital Hall effects being controlled by the same non-Abelian Berry curvature. Our derivation yields a rigorous and general framework for unambiguously determining the physical current density of non‑conserved quantities. Except in the rare instances of orbital degeneracy extending beyond Kramers’ theorem, the intrinsic orbital Hall effect is shown — contrary to widespread assertions — to arise from spin-orbit interaction, at least as a second‑order effect.   Furthermore, we uncover an interband mechanism whereby gradients in the longitudinal linear response to an electric field — ubiquitous near conductor boundaries and interfaces — can locally generate angular momentum density. This mechanism is expected to dominate orbital edge accumulation in light non‑magnetic transition metals. In the case of a simple isotropic metal model, this effect can be understood as a coupling between orbital angular momentum and charge‑current vorticity mediated by the quantum geometry. The explicit expressions derived from this model reproduce the qualitative features of recent orbital edge accumulation observations, while predicting a confinement to within a few electron mean free paths from the boundaries, which seems also compatible with experimental findings. We shall stress that while we have adopted an atomic center definition of the orbital angular momentum operator in the treatment of this simple model, none of our general results depends on this. Indeed, the alternate definition of the electron orbital angular momentum \cite{cysne:2022,pezo:2022,busch:2023}, in the spirit of the modern theory of orbital magnetization \cite{thonhauser:2005,xiao:2005,shi:2007}, also results in an operator becoming purely interband for non-magnetic centrosymmetric metals with Kramers´ twofold degenerate bands in the vanishing spin-orbit interaction limit \cite{supp}, hence leading to all the same conclusions. We also want to outline that the newly identified  effect generating angular momentum edge accumulation is entirely distinct from quantum interference effects previously considered by other authors \cite{tokatly:2010, busch:2023, voss:2025}. While mandating a major upheaval in the interpretation of recent orbital edge accumulation experimental observations, our framework more broadly opens a new chapter in the theoretical approach to orbitronics, in which most previous computations of the intrinsic orbital Hall effect conductivity and related transport coefficients, {\em e.g.}, orbital diffusion tensors \cite{ning:2025}, shall be done anew with the proper definition of orbital current. 

Finally, looking beyond orbitronics {\em per se}, we emphasize that our work provides the theoretical foundation for a new era of {\em electron‑flow gradient engineering} aimed at electric magnetization control. In particular, Eq.~(\ref{eq:x_int2}) can be straightforwardly specialized by lifting the assumption of time‑reversal invariance, with all interband matrix elements reducing to scalars connecting non‑degenerate bands. This yields a general expression for the spin and orbital angular momenta densities locally induced by current gradients in magnetically ordered regions, which will typically include non‑collinear components. Accordingly, Eq.~(\ref{eq:x_int2}) offers, for the first time, a rigorous framework and first‑principles formulation for the quantitative modeling of current-gradient‑induced torques in heterostructures, generalizing the previously introduced concept of spin-vorticity coupling \cite{matsuo:2013,matsuo:2017,takahashi:2016,kobayashi:2017}, and opening a new path toward their systematic optimization. This new perspective should motivate a re‑examination of recent experiments on metal \cite{nakayama:2023,nakayama:2025,yi:2025}, metal–semiconductor \cite{horaguchi:2025}, and metal–oxide \cite{gao:2025} heterostructures, even if in some cases one must also consider the possible role of {\em intraband} diffusion of orbital angular momentum generated by orbital Edelstein effects at interfaces \cite{krishnia:2023,nikolaev:2024,krishnia:2024}.
\vspace{0 pt}

\section*{End Matter}

On the matter of the ubiquitous transverse gradients in the longitudinal intraband response to an electric field that appear near conductor edges, we emphasize in the main text the role of atomic‑level boundary roughness and disorder as the primary mechanism. However, we shall also stress than even atomically flat and ordered boundaries can also induce such gradients, due to the complex angular dependence of their quantum‑mechanical scattering matrices. 

On the matter of our choice of a minimal two‑dimensional, two‑band model to illustrate with a concrete calculation the newly uncovered interband edge accumulation effect, we want to emphasize its universal relevance despite its simplicity. This model captures the essential physics in a way that is robust and independent of parameter choice, even though it cannot reproduce the full complexity of real (often three‑dimensional) non‑magnetic centrosymmetric metals. The key feature is the orbital‑composition dependence of Bloch states on quasi‑momentum within the Brillouin zone — a universal property of crystalline solids — which suffices for the emergence of the effect under discussion, as it essentially ensures a non‑vanishing interband Berry connection. 

On the matter of the observability of the newly uncovered interband effect, it is beyond the scope of the present work to develop a microscopic theory of light–matter interaction in the presence of the predicted out‑of‑equilibrium orbital angular momentum density. Nevertheless, we emphasize that any nonzero electronic angular momentum density, irrespective of its microscopic origin, constitutes a pseudovector order parameter which, in linear response to an optical electric field, permits antisymmetric contributions to the dielectric tensor—terms otherwise forbidden in equilibrium for a non‑magnetic centrosymmetric metal. It should therefore be expected that this effect manifests in reflection ellipsometry measurements on metallic systems as a current‑induced gyrotropic signal near the edges. By the same symmetry argument, provided the additional conditions for luminescence are satisfied, circularly polarized electroluminescence or photoluminescence should arise in suitably engineered semiconductor samples.

\begin{acknowledgments}
\vspace{-5 pt}
We thank Albert Fert, Mairbek Chshiev, Gerrit E.W. Bauer, Paul J. Kelly, Dimitrie Culcer, Mikhail Titov, Armando Pezo, Jing Li and Xavier Waintal for fruitful discussions. This research has been supported by the EIC Pathfinder OPEN Grant No. 101129641 “OBELIX” and a France 2030 government grant managed by the French National Research Agency PEPR SPIN Grant No. ANR-22-EXSP0009 (SPINTHEORY). 
\end{acknowledgments}

\bibliography{berry,berry2}

\end{document}